\documentclass{iopart}
\usepackage{amsmath}
\usepackage{graphicx}
\usepackage{iopams}
\usepackage{url} 

\begin{document}
\title[Precision studies of short-range forces employing
prototypes of GWI]%
{Precision studies of Casimir force and short-range gravity employing
prototypes of interferometric gravitational wave detectors }

\author{ G. Rajalakshmi and C. S. Unnikrishnan}
\address{ Gravitation Group, Tata Institute of Fundamental Research, Mumbai -
400 005, India }

\begin{abstract}
We discuss experimental schemes to measure the Casimir force and short range forces from hypothetical modified gravity with unprecedented sensitivity using highly sensitive prototype gravitational wave detectors as displacement sensors. The finite temperature  effects of  the Casimir force would be detectable with a sensitivity better than $1\%$ for separation exceeding 30 microns. Constraints on short range modifications to gravity can be improved in the distance range of 10-100 microns. 
\end{abstract}

\maketitle

\section{Introduction}
In the last decade, experimental investigations of the Casimir force \cite{Lamoreaux97,Mohi98,Chan01,Ruoso02,Decca03E,Decca2003,Chen05,Decca2007,Raji-T,MG11} have been taken up with renewed interest especially because of its significance to Cosmology and grand unification theories \cite{Bordag01,Milton2002,Rey2002,Adel2003,Milton2004, Lamor2005, Mosto2006, Unni2008}. Considerable attention has also been devoted to theoretical investigations of Casimir force that take into account realistic situations of experiments \cite{Reynaud2000,Bordag2000,Mosto2006, Milton2008,Mosto2009}. More
recently, lateral Casimir force in corrugated structures has also been
measured \cite{Chiu2009,Chiu2010}. However, the large finite temperature effect of the Casimir force remains elusive in experiments so far due to their lack of
sensitivity in the distance range where the effect is significant, beyond
about 5 microns. In contrast, the Casimir-Polder force between atoms and a
surface has been measured including the finite temperature effects
remarkably well recently \cite{Cornell2007}. Accurate theoretical/numerical
calculations of the Casimir force are essential to be able to compare with
experiments and look for new forces in the sub-micron regime. Diverse
experimental apparatus ranging from Atomic Force Microscope (AFM) to
Micro-Electro-Mechanical Systems (MEMS) and highly sensitive torsional
pendulums have been used in these experiments.

Extensions to standard model of particle physics, predict the existence of
new particles that mediate new forces. String and M-theories that attempt to
unify the fundamental forces close to the Planck scale or the theories with
large extra dimensions that attempt unification of the fundamental forces at
the TeV scale of electro-weak symmetry breaking, predict variations from the
inverse square law of gravity at sub-millimetre distances. A review of the
experimental and the theoretical status of the inverse square law tests can
be found in ~\cite{Adel2003,Adel2005,Newman2009}. These deviations are
usually parametrized by the addition of a Yukawa-type correction term to the
Newtonian potential. Thus, 
\begin{equation}
U(z)=-\frac{GM}{z}\left( 1+\alpha e^{-\frac{z}{\lambda }}\right)
\label{Pot-Yukawa}
\end{equation}%
where $G$ is the Newtonian gravitational constant, $M$ the mass and $z$ is
the distance in three dimensional space, $\alpha $ represents the coupling
strength of the new interaction and $\lambda $ its range. One of the more
popular extra-dimensional theories is the Randall-Sundrum(RS) brane-world
model with 5-dimensions \cite{Randall1999-1,Randall1999-2}. In this model,
the corrected potential is given by, 
\begin{equation}
U_{RS}(z)\approx \dfrac{GM}{z}\left( 1+\dfrac{l_{s}^{2}}{z^{2}}\right)
\end{equation}%
where $l_{s}$ is the Randall-Sundrum parameter. Deviations at distance
scales larger than millimeter have been ruled out by astrophysical bounds
and laboratory experiments. In the sub-mm range the present constraints are
less stringent. A laboratory measurement of gravitational force in this
range is the best way to place bounds on the predictions of the new
theories. Measurements of the Casimir force provides the best constraints in the
sub micron distance regime. Compared to these forces of modified gravity,
the Casimir force is much larger, given by 
\begin{equation}
F_{C}=\frac{\pi ^{2}\hbar c}{240z^{4}}\simeq \frac{10^{-7}}{z(\mu m)^{4}}~%
\text{Newtons}
\end{equation}%
Therefore, careful subtraction or shielding of the Casimir force is required
to arrive at constraints on modified gravity at a scale below $100~\mu$m.

Another field of experimental gravitation that has progressed significantly
in the last two decades is that of Gravitational Wave detectors.
Interferometric detectors have reached sensitivity levels for displacements
of the order of $10^{-19}m/\sqrt{Hz}$ , limited only by shot noise \cite%
{LIGO2009,Virgo2008,Geo2008} in the frequency band above 100 Hz or so. The
proposed advanced detectors will try to beat that limit using quantum
squeezing techniques \cite{Vahlbruch2008,Goda2008,Khalili2009}. In low
frequency region ($<100$ Hz), the sensitivity is limited by radiation
pressure noise and seismic noise. The high frequency limitation is photon
shot noise that scales as $1/\sqrt{N}.$ The detectors are Michelson-type
interferometers with Fabry-Perot cavities in the two arms to amplify the
tiny relative displacement of the suspended mirrors. Including the
Fabry-Perot enhancement due to the number of foldings equal to the finesse $%
F_{FP}$ and the reduction in shot noise due to power recycling with finesse $%
F_{PR}$ we can write the shot noise limited displacement sensitivity for an incident  light power $P$, as 
\begin{equation}
\delta x\simeq \frac{\lambda /2}{\sqrt{P/h\nu }}\frac{1}{F_{FP}}\frac{1}{%
\sqrt{F_{PR}}}
\end{equation}

\begin{figure}
\includegraphics[width=0.9\columnwidth]{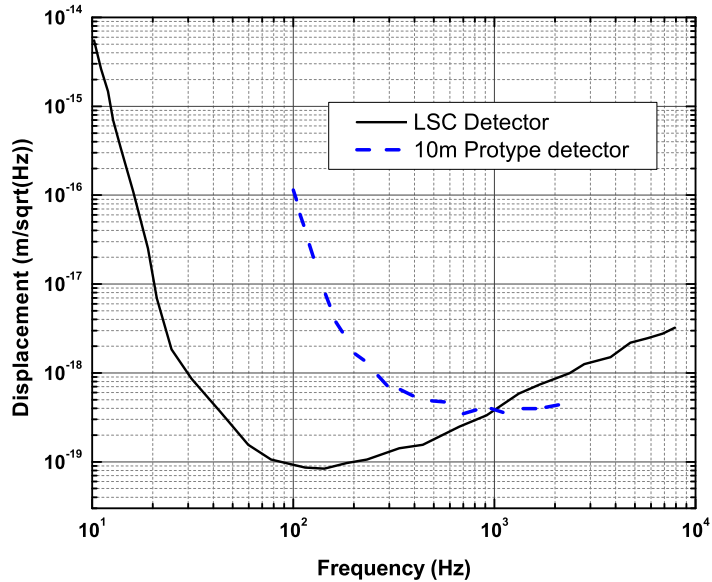}
\caption{Typical displacement sensitivity of long baseline Gravity Wave
detectors, \protect\cite{Geo2008,Clubley2001}}
\label{GWI-sensitivity}
\end{figure}

	The displacement sensitivity of a typical gravitational wave
interferometer (GWI) is shown in Fig.[\ref{GWI-sensitivity}]. 
While the sensitivity to the strain, depends on the arm length of the
interferometric detector, the displacement sensitivity is determined by the
basic optical and noise-isolation design and even a prototype gravitational
wave detector with short arm length of a few meters is capable of
measurements of displacement with a sensitivity below $10^{-18}m/\sqrt{Hz}$ %
\cite{Clubley2001}. (Fig.[\ref{GWI-sensitivity}]). They are ideally suited
for unprecedented high precision measurements of short range forces that can
be modulated at frequencies above 100 Hz or so for an extended period, as in
the case of the Casimir force. With phase sensitive integration of the
signal for a few hours, a displacement signal of $10^{-20}$~m can be pulled
above the shot-noise. This is the basis of our proposed experiment to
measure Casimir force and gravity in the sub-mm range using such a detectors
as the sensitive force transducer.

\section{The Experimental proposal}

Prototype interferometer detectors have suspended end mirrors that are a few
kilograms in weight, determined by considerations of cavity losses, thermal
noise and stability of suspension. A static force of $F$ would
result in a static angular deflection of $\theta =F/mg$ and a static
displacement of $\delta l_{s}\simeq \theta l=Fl/mg,$ where $l$ is the length
of the mirror suspension. The zero-temperature Casimir force, for example,
at the relatively large separation of $100~\mu$m, where no measurement have
been possible, or even foreseen, is about $10^{-15}$~N for $1$~cm$^{2}$ of
surface and the static displacement of $\delta l_{s}\simeq 10^{-17}$ m, for
a 5~Kg mirror on $0.5$ m suspension is well above the sensitivity of the
interferometric detector. The finite temperature effect is even larger,
about $4\times 10^{-14}$ N. However, near zero frequency the interferometer
detector is noisy and its useful sensitivity starts from about 100 Hz or so.
If the force is modulated at high frequency, the response of the mirror
decreases as $\delta l_{s}/\omega ^{2},$ but the sensitivity remains
sufficiently high for measurements of the modulated force with very good
precision in a range of hitherto unexplored separation, $5~\mu m-30~\mu m$.
With longer integration, the measurement can be extended to $100~\mu$m.
Such a measurement is significant and important on two counts, (a) it covers
a range of distances that has never been explored in any previous
measurements and will remain outside the scope of earlier techniques, (b)
the finite temperature effect will be detected and studied in great detail
for the first time.

\begin{figure}
\begin{center}
\includegraphics[height=4.7cm]{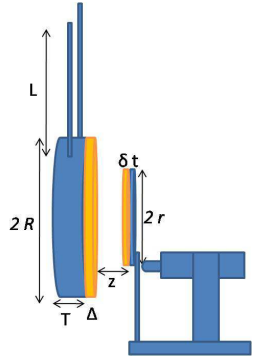}
\caption{Experimental Schematic: One coated plate of large radius is freely suspended while the other coated plate of smaller radius is fixed firmly at the tip of a cantilever spring. The spring is pushed by piezo-electric actuators in order to change the separation between the two plates.   }
\end{center}
\label{expt-scheme}
\end{figure}

The experiment we propose is to measure the force of attraction between one
of the suspended mirrors of the GWI and another movable mirror/plate of
smaller diameter (which we call test plate) that is fixed close to it. The
separation between the mirror and the test plate can be adjusted using
piezo-electric actuators. Fig.[\ref{expt-scheme}] shows a schematic of the
experimental arrangement. This arrangement would require that the
interferometer mirror be coated on both sides, alternatively, a coated lip
can be attached to the suspended mirror. In either of the configuration, the
equilibrium position of the suspended mirror would be affected by the force
of attraction between the suspended mirror and the test plate. When the test
plate is modulated about a fixed position, the suspended mirror would be
displaced and this would constitute a signal in the GWI. In the following
sections we will model the force acting on the suspended mirror and derive
the expected displacement. 

\section{Forces on the suspended mirror}

\subsection{Casimir force between coated parallel plates}

The Casimir force per unit area between parallel metal surfaces at absolute zero
temperature is, 
\begin{equation}
F_c(z) = -\frac{\pi^2 \hbar c}{240z^4}.
\end{equation}
At any finite temperature, the force per unit area is given by, 
\begin{align}
F_{c}^T(z) & = -\frac{k_{B} T}{4\pi z^{3}}\sum_{n=0}^{n}\hspace*{-1.0mm}{}^{%
\prime}\int_{nx}^{\infty}\frac{dyy^{2}}{e^{y} - 1} \quad \mathrm{where}\; \;
x \equiv4\pi k_{B} Tz/\hbar c  \label{FcasT} \\
F_c^T(z) & \simeq -\frac{\zeta(3) k_BT}{4\pi z^3}\quad \text{at\;high\;}%
T\;(i.e.\;x\gg1) \\
& \quad \quad \mathrm{with\ }\zeta(3)= 1.20206  \nonumber
\end{align}

Thus the distance dependence of the force changes from $z^{-4}$ for zero
temperature to $z^{-3}$ in the case of finite temperature. The important
non-dimensional parameter, that distinguishes the domains of \textit{high}
and \textit{low} temperature, is $x=4\pi k_{B}T z/\hbar c.$ The finite
temperature effect becomes dominant for separations $z$, greater than the
thermal wavelength $\lambda_T \simeq \hbar c / k_{B}T$. Detailed
calculations of the finite temperature effects for various configurations
have been presented in \cite{Reynaud2000,Bordag2000}. These calculations
also include corrections due to finite conductivity and surface roughness.
All details on the calculation of the Casimir force between real material
bodies can be found in the recent monograph \cite{Bordag2009}.

In the proposed experiment, the mirror and the test plate would be at
temperatures close to about 300~K, finite temperature corrections become
appreciable beyond about $3~\mu$m. Hence, in the distance range of $10-100~\mu
$m where we propose to do the experiment, the Casimir force on the suspended
mirror due to the test plate would be given by Eqn.\ref{FcasT}. Correction
due to the finite conductivity and surface roughness is negligibly small over the range of the proposed experiment. As shown in \cite{Bordag2000} for gold coatings, the correction due to finite conductivity is about $0.66\%$ at $10~\mu$m separation and scales down to $0.066\%$ at $100~\mu$m separation.  Also in this distance range, Casimir force will be the strongest force acting on the suspended mirror.

\begin{figure}
\begin{center}
\includegraphics[height=4.7cm]{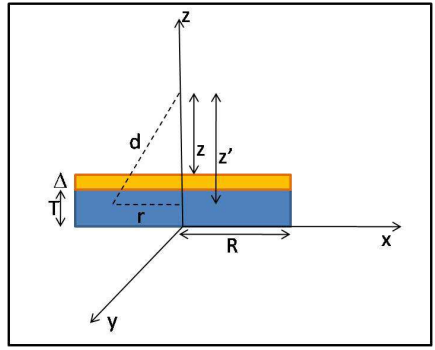}
\caption{The reference axis used in the derivation of the force expressions are shown in relation to the dimensions of the large plate. The suspension axis would be along the x-axis.}
\label{reference-axis}
\end{center}
\end{figure}

\subsection{Gravitational force between coated parallel plates}

Next to Casimir force, gravitational interaction between the plates will be
the most important. If the suspended mirror is of radius $R$
and thickness $T$ and the test plate is of radius $r$ and thickness $t$,
such that $T>>t$, then in the simplest case, the gravitational force between
the plates will be independent of the distance between them and would depend only
on their respective densities $\rho_1, \rho_2$ and thickness. To the leading
order in T and t, it is given by, 
\begin{equation}
F_{grav}(z) \approx 2 \pi^2 r^2 G \rho_1 \rho_2 Tt
\end{equation}
The mirrors are typically made of glass substrate coated with  metal or
dielectric layers whose densities are very different from the substrate.
Consider such a mirror of radius $R$ with substrate density $\rho_{sub}$
and thickness $T$ and coating density $\rho_{coat}$ and thickness $\Delta$,
the gravitational potential due to the plate at a distance, $z$ (Fig.~[\ref{reference-axis}]) from the surface of the plate is obtained by integrating the
contribution due to various mass elements on the mirror, 
\begin{align}
U_{grav}(z) = & - G \int \frac{\rho(z)}{\sqrt{r^2+z^2}}. 2\pi r dr dz \\
= & - 2 \pi G \int_{z+T+\Delta}^{z} dz \ \rho(z) \int_{0}^{R} \frac{dr \ r}{\sqrt{%
r^2+z^2}} \\
=& -2 \pi G \int_{z+T+\Delta}^{z} dz \ \rho(z) \left[ \sqrt{R^2 + z^2} - z %
\right]
\end{align}
For $R >> Z$, the potential is given by, 
\begin{align}
U_{grav}(z)\simeq & - 2 \pi G R (\rho_{sub} T + \rho_{coat}\Delta) 
-\frac{\pi G}{R}\frac{\rho_{sub}}{3} \left [(z + \Delta)^3 -(z + \Delta +
T) ^3 \right ]  \\
& -\frac{\pi G}{R}\frac{\rho_{coat}}{3} \left [z^3 -(z + \Delta) ^3 \right ] 
 - 2 \pi G \left[ \rho_{sub} \left\lbrace T^2 + 2(z +
\Delta)T \right\rbrace + \rho_{coat} (\Delta^2 + 2 z \Delta) \right]  \notag
\end{align}

The force due to this potential at the point z is, 
\begin{align}
f_{grav}(z) = & - \frac{\partial U}{\partial z} \\
\simeq & - 2 \frac{\pi G}{R}\left( \rho_{sub} z T + \rho_{coat} z \Delta
\right) + \pi G \rho_{sub} \left( 2 T - \frac{2\Delta T + T^2}{R} \right)  \notag
\\
& + \pi G \rho_{coat} \left( 2 \Delta - \frac{\Delta^2}{R} \right)
\end{align}

The gravitational force between this and the test plate of radius $r$ with
substrate density $\rho_{sub}$ and thickness $t$ and coating density $%
\rho_{coat}$ and thickness $\delta$, placed at a distance $z$ from the
mirror surface would be got by integrating this force over the volume of the
test plate. Thus the gravitational force between the plates will be given
by, 
\begin{align}
F_{grav}(z)  \notag \\
= & \int f_{grav}(z^{\prime}) \rho(z^{\prime}) dV \\
= & \int_{z}^{z+t+\delta} dz f_{grav}(z^{\prime}) \rho(z^{\prime})
\int_{0}^{r} 2 \pi r dr \\
\simeq &\ \pi G r^2 \Bigg[ \left( \rho_{sub} t + \rho_{coat} \delta \right)
\cdot %
\left\lbrace \rho_{sub} \left(2T -\frac{2 \Delta T -T^2}{R} \right) +
\rho_{coat}\left( 2\Delta - \frac{\Delta ^2}{R} \right) \right\rbrace  \notag
\\
& + \left( \rho_{sub} T + \rho_{coat} \Delta \right) \cdot  
 \left\lbrace \frac{\rho_{sub}\left(t^2-2(z+\delta )t\right) +
\rho_{coat}\left(\delta ^2 - 2 \delta z\right)}{R} \right\rbrace \Bigg ] 
\end{align}
As expected for an inverse square law, the gravitational force is largely
independent of the distance between the plates for $R >> z, r $. The
dependence on $z$ is largely due to edge effects and the leading
contribution goes as $z/R$.

\subsection{Force due to Yukawa-type deviations to gravity}

The Yukawa-type correction to Newtonian gravity will lead to a force which
will fall exponentially as the range of the interaction $\lambda$, the
coupling strength would be modified by a factor $\alpha$, the coupling
constant of the `new force'. Thus, in the simplest case, the force between
two plates would be, 
\begin{equation}
F_{Yuk}(z) = 2 \pi^2 r^2 G \alpha \lambda^2 e^{\frac{-z}{\lambda}} \left[
\rho_1 \left( 1 + e^{\frac{-T}{\lambda}}\right) \right] \left[ \rho_1 \left(
1 + e^{\frac{-t}{\lambda}}\right) \right]
\end{equation}
In the presence of coating, this force can be derived from the potential
following the same procedure as that of the gravitational force. Thus, 
\begin{align}
F_{Yuk}(z)  = 2 \pi ^2 r^{2 } G \ \alpha \lambda ^2 e^{-z/\lambda}& \Big[ %
\rho_{sub}\ e^{-\Delta /\lambda }\left(1 -\ e^{-T/\lambda} \right)  
+ \rho_{coat}\left(1 -\ e^{-\Delta /\lambda}\right)\Big]\notag \\
&\Big[\rho_{coat} \left(1- e^{-\delta /\lambda }\right) 
 + \rho_{sub} e^{-\delta /\lambda } \left(1- e^{-t/\lambda }\right) \Big]
\end{align}

\subsection{Force due to RS-type correction to gravity}

The RS-type modification to gravity will give rise to a force that is
diminished by $l_s^2$ as compared to gravity and has a slow logarithmic
dependence on the separation. The RS-correction term for uncoated plates
will be,

\begin{equation}
F_{RS}(z) = 2 \pi^2 r^2 G l_s^2 . \rho_1 \ln \left[ \frac{z+T+t}{z+T}\right]%
. \rho_2 \ln \left[ \frac{z+t}{z}\right]
\end{equation}

For experimental mirrors with coated surfaces, the force can be shown to be, 
\begin{align}
F_{RS}(z)= -2 \pi^2 r^{2 }G l_s^2 
&\Bigg[ \rho_{coat}^2 \left\lbrace \ln \left(\frac{z + \delta + \Delta}{z +
\Delta }\right) -\ln \left( \frac{z + \Delta }{z}\right) \right\rbrace 
\notag \\
& + \rho_{sub} \rho_{coat} \left\lbrace \ln \left(\frac{z +
\delta +T+ \Delta }{z + T+ \Delta } \right) -\ln\left( \frac{z +\delta +
\Delta }{z+ \Delta }\right) \right.   \\
& + \left. \ln \left(\frac{z +t+ \delta + \Delta }{z +\delta + \Delta }%
\right) -\ln \left( \frac{z +t+\delta}{z+ \delta}\right) \right\rbrace 
\notag \\
& +\rho_{sub}^2\left\lbrace \log\left[\frac{z +t+ \delta
+T+ \Delta }{z +\delta + T+ \Delta } \right]-\log\left[ \frac{z +t+\delta +
\Delta }{z+\delta + \Delta }\right]\right\rbrace\Bigg] \notag
\end{align}

\section{Expected signals}

\begin{figure}
\begin{center}
\includegraphics[width=\columnwidth]{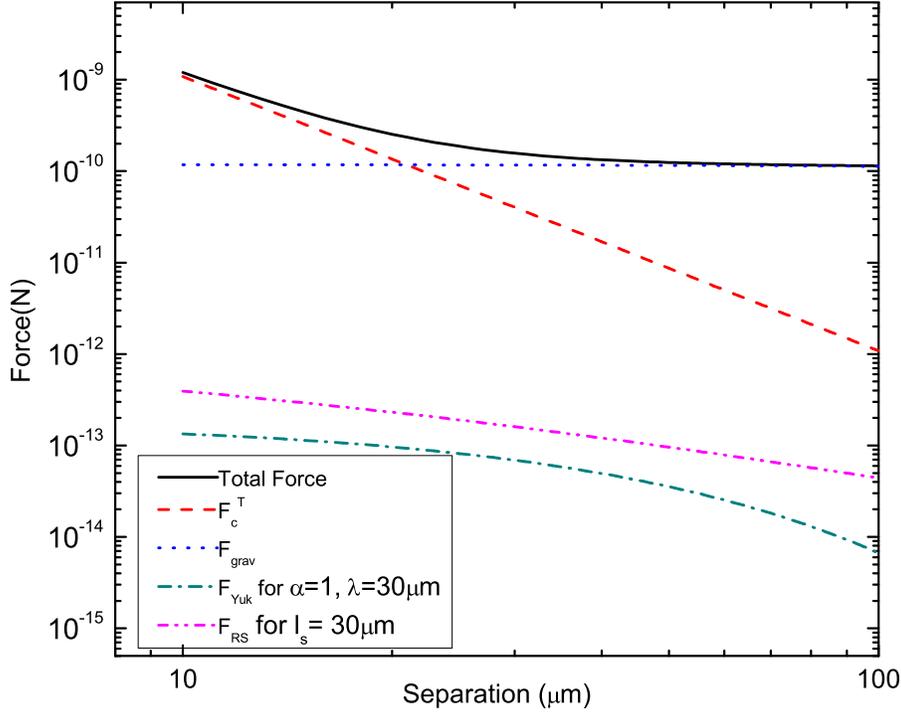}
\caption{Forces acting on the Interferometer Mirror}
\end{center}
\label{forces}
\end{figure}
\begin{figure}
\includegraphics[width=\columnwidth]{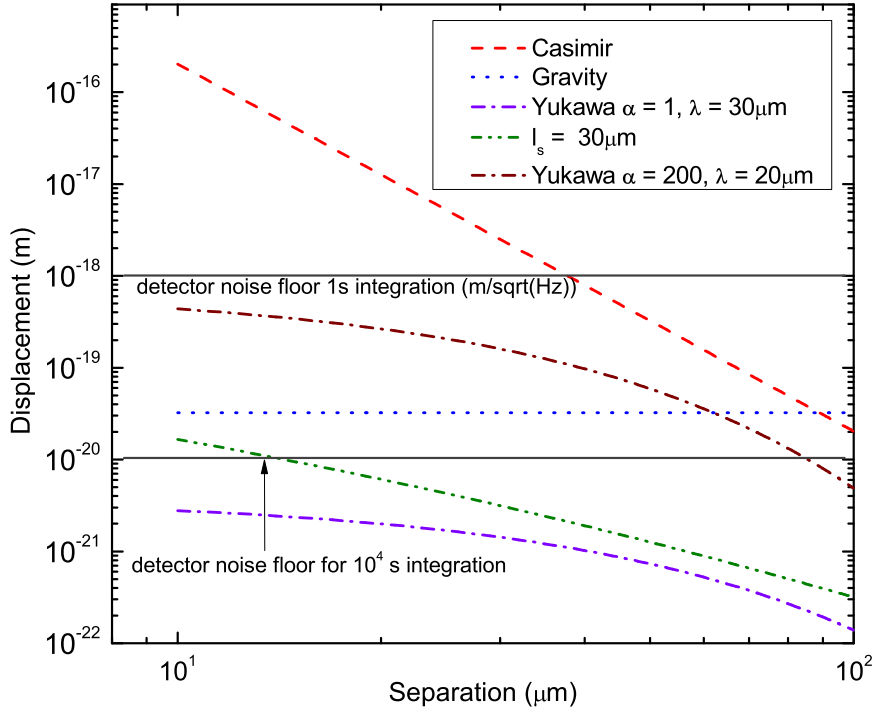}
\caption{Expected displacement of the mirror for $2\protect\mu $m modulation
at $200$Hz}
\label{displacements}
\end{figure}

The forces described above are plotted in Fig.~[\ref{forces}] when the
suspended mirror has a radius $R=10~$cm and is made of glass of thickness $%
T=5~$cm, coated with gold layer of thickness $\Delta =30~\mu $m and the test
mirror has a radius $r=3~$cm and is made of glass of thickness $t=0.1$~mm,
coated with gold layer of thickness $\delta =30~\mu$m. The displacement of
the mirror due to this force would be $(Force\times L/m\ g)$. GW interferometers are designed to look at changes in the mirror position; however their noise floor at DC and low frequencies is too large for the signal to be measurable. Thus, in
order to create a measurable signal, the separation between the plates will
be modulated about a fixed separation. The displacement measured will be
proportional to the spatial derivative of the force and the amplitude of the
modulation. When we modulate at a frequency $\nu $, far from the resonance
of the detector, the amplitude of the signal is suppressed by $1/\nu^{2}$. Fig.~[\ref{displacements}] shows the expected displacement of a mirror
of mass $m=10$~Kg, with a suspension length $L=0.5$~m due to the various
forces for a modulation amplitude of $2~\mu$m at $200$~Hz. The dominant
signal is due to the Casimir force with expected displacements at $10~\mu $m
separation between the plates that are 10 times the sensitivity of the GWI.
Even though the gravitational force is stronger than Casimir force for
separations larger than $\sim 20~\mu$m, it does not contribute to the signal
even at separations of $\sim 90~\mu$m as it depends very weakly on $z$. The
signal to noise ratio will increase by a factor of 100 if we integrate the
signal for $10^{4}$~s. Hence, even at separations as large as $30~\mu$m the
Casimir force can be measured to an accuracy of better than $1\%$. This
would be the most accurate measurement of the Casimir force so far at separations greater than $1~\mu$m. It might be possible to improve this even further if the useful frequency band of the detector can be brought down to 100 Hz, which is feasible with low frequency isolation techniques \cite{AIGO2006, AIGO2010}. 

Earlier measurements of Casimir force at separations larger than $1~\mu$m were
plagued by electrostatic forces arising  due to patch fields, contact potentials and static charges. As documented by every experimenter trying to measure the
Casimir force at micron scale separations, the electrostatic forces have to be 
carefully measured and subtracted to reveal the presence of Casimir force. Some of the serious concerns regarding this issue have been expressed by Speake et. al.~\cite{Speake2003}, who derive an expression for the dependence of the  force due to random variation of the electrostatic potential, on the separation between the plates. Decca et. al. \cite{Decca2005} have also shown that the correction due to these effects can be as small as $0.037\%$ even at separations as small as $160$~nm and already reduces to $0.027\%$ at $170$~nm separation. Thus, for separations larger than $10~\mu$m and for the configuration of the plates in the proposed experiment, the gradient of the stray electrostatic forces will be small and hence its contribution to the signal will be atmost at the level of the gravitational force. We expect that patch field effects can be measured and corrected for at the required level at the relatively large separations we plan to make the measurements. At separations beyond $10~\mu$m, the mirror can be parallel transported once it is aligned parallel by fixing it on a dual flexure stage. The  parallelism can also be monitored interferometerically.

\begin{figure}
\includegraphics[width=\columnwidth]{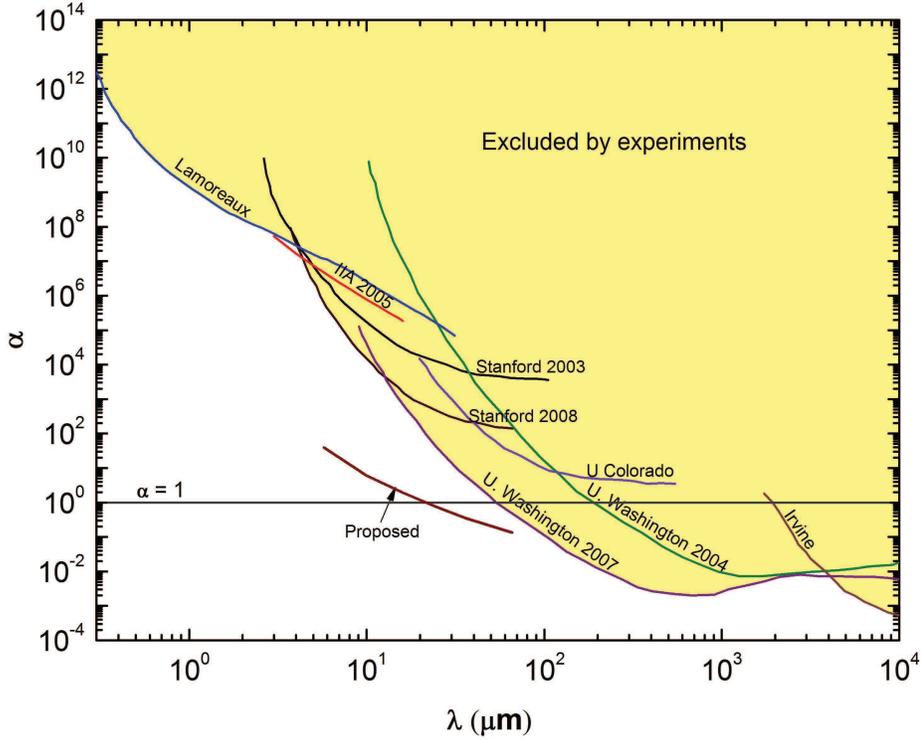}
\caption{Constraints on the inverse square law violating Yukawa interactions
derived from experiments as indicated in Ref.~\cite%
{Lamoreaux97,Raji-T,MG11,Stanford2003,Stanford2008,Hoyle2004,Kapner2007,Long2003,Irvine1985}
}
\label{constraints}
\end{figure}

The parameters of predicted inverse square law violating interactions can be
constrained by comparing the measured force with the theoretically expected
force. A $1\%$ measurement of Casimir forces in the $10~\mu$m to $60~\mu$m
range would place constraints of order $1\times 10^{-13}~$N, corresponding to
an $\alpha $ of $1$ for $\lambda =30~\mu$m on the corrections to inverse
square law. The existing limits of the violation parameters derived from
experiments are shown in Fig.\ref{constraints}. The parameter space above
the curves is excluded by experiments. The existing constraints on Yukawa
type interaction in the range of our experiment is indicated by $\alpha \leq
200$ at $\lambda =20~\mu$m. Our proposed experiment will place limits in the
level of $\alpha =1$ at $\lambda =30~\mu$m, and study the inverse square law
in the range $10-30~\mu$m.  This range of the parameters $(\alpha ,\lambda )$
has so far not been probed with sensitivity sufficient to significantly
constrain particle physics models with implications to the gravitational
interaction.  By mounting a thin conducting membrane between the two mirrors
at a fixed distance, say $5~\mu$m from the suspended mirror, the Casimir
force between the mirrors can be kept constant while varying the distance to
the movable mirror to measure the inverse-square dependence of gravity. By
modulating the plate at $200$~Hz and integrating for $10^{4}$~sec, Yukawa type
interaction with $\alpha =200;\lambda =30~\mu$m would give rise to signals
that can be detected to about $5\%$ indicating significant improvement
over previous measurements ( Fig.~[\ref{displacements}]).

The best constraints on $\alpha $ in the range of $\lambda $ below $10~\mu$m
are from measurements of Casimir force \cite{Lamoreaux97,Raji-T}. To improve
these constraints it is necessary to measure Casimir force in this distance
range with accuracies better than $0.1\%$. There is considerable practical
difficulty in measurements in this distance range with flat plates since
maintaining parallelism of relatively large plates to much less than $1\%$
of their separation, viz., $1-10~\mu$m, is a difficult task.
This difficulty can be avoided by replacing the test plate by a convex surface of
large radius of curvature, at the cost of the displacement signal. The
forces between plates in this scheme can be derived accurately enough for
comparison with precision experiments by applying the proximity force
approximation \cite{Derjaguin60}. Expressions for Casimir force, Gravity and
Yukawa type correction to gravity for this geometry are derived in \cite%
{Raji-T, Reynaud2009, Decca2009}. 

\section{Conclusion}
We have presented an experimental scheme employing high sensitivity
prototype GW interferometer detectors to measure the Casimir force at
separations of $10-100~\mu$m with unprecedented accuracy. Finite
temperature effects can be detected and explored in detail by studying the
force law as a function of separation and by making measurement at various
temperatures between room temperature of about $25^{\circ }$C and $%
100^{\circ }$C. Other aspects of the Casimir force, including peculiarities
arising with corrugated surfaces that break translational symmetry
\cite{Chiu2009, Reynaud2008} are expected to be accessible with better
sensitivity in our scheme. An experiment to search for hypothetical 
modifications to inverse square law of gravity in the $10~\mu$m to 
$100~\mu$m can be performed and improved limits
can be placed on the parameters of inverse square law violating
interactions. These experiments will be taken up in our proposed 3-meter
interferometer in the Indian gravitational wave research initiative \cite%
{Indigowebsite}.
\section*{Bibliography}
\bibliographystyle{unsrt}
\bibliography{gw-cas-references-revised}

\end{document}